\documentclass[12pt]{article}

\usepackage{graphicx}
\newcommand{\ol}[1]{\overline{#1}}

\newcommand{\ep}{\varepsilon}
\let\hat = \widehat
\let\tilde = \widetilde
\newcommand{\beq}{\begin{equation}}
\newcommand{\eeq}{\end{equation}}
\newcommand{\beqa}{\begin{eqnarray}}
\newcommand{\eeqa}{\end{eqnarray}}
\newcommand{\rec}[1]{\mbox{$\frac{1}{#1}$}}
\newcommand{\half}{\rec{2}}
\newcommand{\mfrac}[2]{\mbox{$\frac{#1}{#2}$}}
\newcommand{\nn}{\nonumber} 

\newcommand{\ee}{{\rm e}} 
\newcommand{\dd}{{\rm d}}
\newcommand{\lbl}[1]{\label{#1}}

\begin{document}

\title{Discrete breathers in honeycomb Fermi-Pasta-Ulam lattices} 

\author{Jonathan AD Wattis and Lauren M James \\
School of Mathematical Sciences, University of Nottingham, \\ 
University Park, Nottingham NG7 2RD, UK. \\ 
{\tt Jonathan.Wattis@nottingham.ac.uk} }

\date{{\footnotesize \today}}

\maketitle

\begin{abstract}
We consider the two-dimensional Fermi-Pasta-Ulam lattice with 
hexagonal honeycomb symmetry, which is a Hamiltonian system 
describing the evolution of a scalar-valued quantity 
subject to nearest neighbour interactions.  
Using multiple-scale analysis we reduce the governing lattice 
equations to a nonlinear Schr\"{o}dinger (NLS) equation coupled 
to a second equation for an accompanying slow mode. 
Two cases in which the latter equation can be solved 
and so the system decoupled are considered in more detail: 
firstly,  in the case of a symmetric potential, 
we derive the form of moving breathers.
We find an ellipticity criterion for the wavenumbers 
of the carrier wave, together with asymptotic estimates for the 
breather energy. The minimum energy threshold depends on the 
wavenumber of the breather. We find that this threshold is 
locally maximised by stationary breathers. 
Secondly, for an asymmetric potential we find stationary breathers, 
which, even with a quadratic nonlinearity generate no second 
harmonic component in the breather. 
Plots of all our findings show clear hexagonal symmetry 
as we would expect from our lattice structure. 
Finally, we compare the properties of stationary 
breathers in the square, triangular and honeycomb lattices. 
\end{abstract}

\section{Introduction}
\label{sec:intro}

Discrete Breathers (DBs) are time-periodic and spatially-localised 
exact solutions which describe the motion of a nonlinear lattice, 
that is, a repeated arrangement of atoms. 
In this paper, we investigate the properties of discrete breathers 
on a two-dimensional honeycomb lattice, seeking conditions 
under which the lattice may support breather solutions. 

The combination of nonlinear interactions and discreteness gives rise 
to breather modes.  The discreteness causes gaps and cutoffs in 
the phonon spectrum, whilst nonlinearity allows larger-amplitude 
waves to have frequencies outside the phonon band. 
MacKay \& Aubry's work \cite{mackay2} established the existence 
of breathers in one- and higher-dimensional lattice systems.  
Flach {\em et al.}\ \cite{fkw94} have shown that 
properties of breathers in the more familiar one-dimensional 
systems apply also to lattices in higher dimensions. 
As well as this analytical work, numerical methods have 
also been applied to DB's in higher dimensional systems.
For example, Takeno \cite{takeno} used lattice Green functions 
to determine approximations to breather solutions in one-, two- 
and three-dimensional lattices.  Burlakov \textit{et al.}\  
\cite{burlakov} found breather solutions numerically 
on a two dimensional square lattice and in \cite{bonart}, 
Bonart \textit{et al.}\ simulated numerically localised excitations 
on one-, two- and three-dimensional scalar lattices.

Although the existence of breathers does not depend on the 
lattice dimension, some of their properties do.
Flach \textit{et al.}\ \cite{flach1} found that minimum energy 
threshold in order to create breathers if the lattice dimensions 
is equal to or greater than a critical value. This threshold energy 
is the positive lower energy bound attained by the breather. 
Strictly, only stationary breathers are necessarily time-periodic; 
however, MacKay and Sepulchre \cite{mackay1} have formulated 
a more precise definition of travelling breathers. 
Moving breathers in two-dimensional lattices were investigated 
in a collection of papers by Marin, Eilbeck and Russell who were 
motivated by the observation of dark lines formed along crystal 
directions in white mica \cite{russel1}.   In mica, potassium atoms 
lie in planes in which they occupy a hexagonal pattern.  Numerical 
simulation of Marin \textit{et al.}\ \cite{marin1} exhibited moving 
breathers which only travelled along lattice directions.  Similar 
results were observed in a further study of Marin \textit{et al.}\  
\cite{marin2}, where two- and three- dimensional lattices of various 
geometries were investigated.  
The mechanical lattice, in which each node can move horizontally and 
vertically is highly complex, and although attempts at a full asymptotic 
analysis have been made (for example, \cite{yi}), the detailed 
understanding of dynamics of such a system is not yet available.  

Currently, there is great interest in the behaviour of honeycomb 
lattices, due to the development of potential applications of graphene. 
For example, Molina and Kivshar \cite{ms} studied the localisation 
and propagation of light along ribbons which have a honeycomb 
structure analogous to graphene.   Bahat-Treidel {\em et al}.\  
\cite{btpsb} have studied the propagation of a field in a photonic 
lattice with Kerr nonlinearity. They show that, in the honeycomb 
lattice, the Kerr nonlinearity produces waves with triangular symmetry. 
Chetverikov {\em et al}.\ \cite{cev} consider a system with 
Lennard-Jones-like interaction potentials.  Using a variety of initial 
conditions, they use numerical simulations, to find outputs which 
bear strong visual similarity with results of bubble chamber experiments. 
A system of spherical particles in a hexagonal structure interacting 
with nearest neighbours via Hertzian contacts is considered by 
Leonard {\em et al.}\ \cite{chong}. They analyse the waves that 
spread through the system following a localised impulse. 
Kevrekedis {\em et al.}\ \cite{kmg} consider interactions which 
include longer-range as well as nearest neighbours in a DNLS model. 
They find that these can stabilise and destabilise solitons. 
Ablowitz and Zhu \cite{ablowitz} use perturbation theory to analyse 
the linear spectrum of a hexagonal lattice near its Dirac point, 
as well as the associated Bloch modes and envelope solutions. 

Herein, we consider an electrical  transmission lattice, in which a scalar 
quantity, for example the charge stored on a nonlinear capacitor is 
defined at each node, with nodes being coupled by linear inductors. 
This paper follows on from from previous work of Butt 
and Wattis \cite{Buttwatt1,Buttwatt2} who studied discrete breathers 
in two-dimensional square and hexagonal electrical lattices. 
In such lattices, the scalar valued functions at each node can be 
thought of as charge, thus there is only one degree of freedom at each 
node. This contrasts with the models simulated by Marin {\em et al.}\  
\cite{marin1,marin2} where there is a vector-valued function at each 
node, the in-plane, horizontal and vertical displacements. 
In \cite{Buttwatt1} the lattice considered has $C_4$ rotational symmetry, 
that is, rotations through any multiple of $\pi/2$ radians maps 
the lattice onto itself.  In \cite{Buttwatt2}, a hexagonal lattice with 
$C_6$ rotational symmetry is analysed, here, a rotation through an 
angle which is a multiple of $\pi/3$ maps the lattice onto itself. 
Although this lattice was formed of tessellating triangles, it is the 
rotational symmetry that gives the hexagonal lattice its name. 
In both cases, the method of multiple scales was applied, leading to 
an approximation for small amplitude breathers and their properties. 
Asymptotic estimates for breather energies were found, confirming the 
existence of minimum threshold energies obtained by Flach \cite{flach1}. 
Numerical simulations showed that there was no restriction on the 
allowed direction of travel.
This result contrasts with the behaviour of the mechanical lattice 
analysed by Marin {\em et al.}\ \cite{marin1}, who find breathers 
which only travel along lattice directions. 

\begin{figure}[ht] 
\includegraphics[width=12cm]{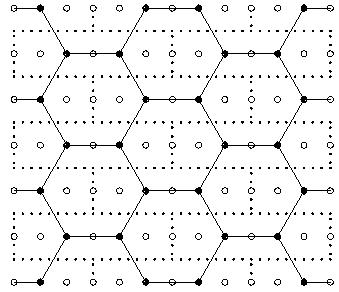}
\caption{The two-dimensional honeycomb lattice. 
Solid circles denote the nodes in the lattice, open circles 
show the unused nodes in the underlying rectangular grid.  
The dotted lines indicate the unit cells, each of which 
contains one left- and one right-facing node.  }
\label{fig:comb}
\end{figure}

This model we consider is simplified, in that only weak nonlinearities 
are considered, and includes no onsite potential. 
We investigate the behaviour of discrete breathers on the  
two-dimensional honeycomb lattice shown in Figure \ref{fig:comb}. 
The lattice possesses $C_3$ rotational symmetry, being 
made up of tessellating hexagons, in which rotation through any angle 
of a multiple of $2\pi/3$ leaves the lattice invariant.  Our aim is to 
investigate the combined leading-order effects of nonlinear 
nearest-neighbour interactions and the honeycomb geometry, 
by finding leading-order asymptotic forms of discrete breathers 
in this lattice. 
This complements previous studies of square and hexagonal lattices 
\cite{Buttwatt1,Buttwatt2}.  Numerical studies of Marin {\em et al.}\  
\cite{marin1,marin2} required the use of an onsite potential as well as 
nonlinear nearest-neighbour interactions to general breathers in 
two-dimensional lattices.  One aim of the current work is to provide 
parameter regimes and initial conditions where breathers may 
exist in a system with only nonlinear nearest-neighbour interactions. 
We follow a similar analytic procedure to that of \cite{Buttwatt1,Buttwatt2}, 
using the method of multiple scales to obtain a system of equations from 
which we derive a nonlinear Schr\"{o}dinger (NLS) equation that allows 
us to determine approximate small amplitude breathers.
However, the analysis of the honeycomb lattice is significantly more 
complicated than the square or hexagonal cases due to the geometry 
of the lattice's interconnections which mean that there are {\em two} 
distinct types of node, which we call left-facing and right-facing. 
The analysis is similar to that of diatomic lattices, as it supports 
two types of mode which can be termed `acoustic' and `optical'. 

In section \ref{sec:model} we derive the governing equations 
and the Hamiltonian structure behind them.   We use the method 
of multiple-scales in Section \ref{sec:gen} to determine 
approximations to small amplitude breathers.  Taking the amplitude, 
$\ep$, as our small parameter, we form a power series expansion, 
equating terms at each order in $\ep$ and each harmonic of a 
fundamental linear mode. A dispersion relation is found in Section 
\ref{subsec:disp},  the plot of which shows some 
of the symmetry properties that we expect to find in the lattice. 
We find, in Section \ref{subsec:nls}, a reduction of the governing 
lattice equations to an NLS equation in two special cases. 
The first case, analysed in Section \ref{subsec:sym}, 
is where the interaction potential is symmetric,  and along with the 
NLS equation, we find an ellipticity criterion for moving breathers; 
this means that only certain combinations of wavenumbers may 
produce a moving breather.  
The second special case, investigated in Section \ref{subsec:asym}, 
covers asymmetric potentials, but is restricted to stationary breathers. 
A relationship between the coefficient of the quadratic and cubic 
nonlinearities is derived in this case. 
We conclude our findings in Section \ref{sec:conc} with a summary 
of the results derived, and suggestions for further study. 

\section{A two-dimensional honeycomb lattice}
\label{sec:model}

\subsection{Geometry of the lattice}
\label{subsec:setup}

We consider the nodes of the hexagonal honeycomb lattice as lying on a 
subset of a rectangular lattice.  First, we introduce orthonormal  
basis vectors $B = \{\textbf{i},\textbf{j}\}$, where $\textbf{i} 
= [1,0]^T $ and $\textbf{j} = [0,1]^T $. The position of the $(m,n)$ 
node of the rectangular lattice is $m\textbf{i} + hn\textbf{j}$, 
with $h=\sqrt{3}$ so that the resulting hexagons are regular. 
In order to specify the honeycomb lattice, we retain only those nodes 
$(m,n)$ for which $m+n$ is an even integer and omit $m=6p+1$, 
$n=$ odd and $m=6p+4, n=$ even.  In Figure \ref{fig:comb} 
the filled circles denote the nodes retained in the honeycomb 
lattice which satisfy these relations, and the open circles show 
all the remaining nodes in the underlying rectangular lattice. 

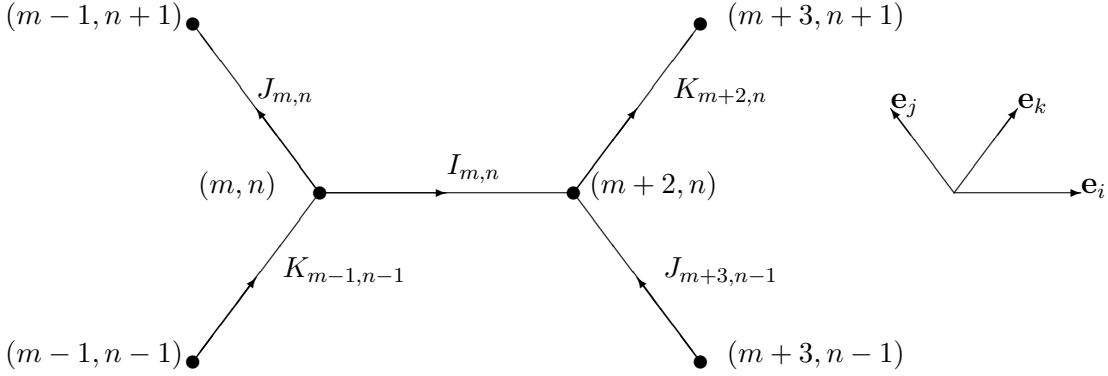
\begin{figure}[ht] 
\begin{picture}(500,200)(-10,0)
\put(116, 64){\circle*{5}}
\put(70, 64){\small $ (m,n)$}
\put(212, 64){\circle*{5}}
\put(218, 64){\small $ (m+2,n)$}
\put(68, 0){\circle*{5}}
\put(-3, 0){\small $ (m-1,n-1)$}
\put(68, 128){\circle*{5}}
\put(-3, 128){\small $ (m-1,n+1)$}
\put(260, 0){\circle*{5}}
\put(270, 0){\small $ (m+3,n-1)$}
\put(260, 128){\circle*{5}}
\put(270, 128){\small $ (m+3,n+1)$}
\put(116, 64){\line(-3, 4){48}}
\put(116, 64){\vector(-3,4){24}}
\put(102,32){\small{$K_{m-1,n-1}$}}
\put(116, 64){\line(-3, -4){48}}
\put(68, 0){\vector(3,4){24}}
\put(246,100){\small{ $K_{m+2,n}$ }}
\put(116, 64){\line(1, 0){96}}
\put(116, 64){\vector(1, 0){48}}
\put(164,71){\small{$I_{m,n}$}}
\put(212, 64){\line(3, 4){48}}
\put(212, 64){\vector(3,4){24}}
\put(92,100){\small{$J_{m,n}$} }
\put(212, 64){\line(3, -4){48}}
\put(260, 0){\vector(-3,4){24}}
\put(246,32){\small{$J_{m+3,n-1}$} }
\put(404, 64){${\bf e}_i$}
\put(380, 96){${\bf e}_k$}
\put(332, 96){${\bf e}_j$}
\put(356, 64){\vector(-3, 4){24}}
\put(356, 64){\vector(3, 4){24}}
\put(356, 64){\vector(1, 0){48}}
\end{picture}
\caption{Enlarged view of the honeycomb lattice }
\label{fig:enlarged}
\end{figure}

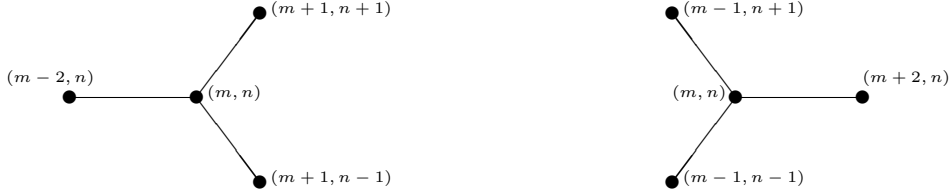
\begin{figure}[ht] 
\begin{picture}(200,120)(0,-20)
\put(96, 64){\circle*{5}}
\put(48, 64){\circle*{5}}
\put(120, 32){\circle*{5}}
\put(120, 96){\circle*{5}}
\put(96, 64){\line(3, 4){24}}
\put(96, 64){\line(3, -4){24}}
\put(96, 64){\line(-1, 0){48}}
\put(100, 64){\tiny $ (m,n)$}
\put(24, 70){\tiny $ (m-2,n)$}
\put(124, 32){\tiny $ (m+1,n-1)$}
\put(124, 96){\tiny $ (m+1,n+1)$}
\put(28,0){\small{(a) Arrangement 1, $\hat Q_{m,n}$ in centre} }
\put(43,-12){\small{neighbouring nodes are $\ol Q_{m-2,n}$, }}
\put(43,-24){\small{$\ol Q_{m+1,n+1}$, and $\ol Q_{m+1,n-1}$.}}
\end{picture}
\begin{picture}(200,120)(0,-20)
\put(96, 64){\circle*{5}}
\put(144, 64){\circle*{5}}
\put(72, 32){\circle*{5}}
\put(72, 96){\circle*{5}}
\put(96, 64){\line(-3, 4){24}}
\put(96, 64){\line(-3, -4){24}}
\put(96, 64){\line(1, 0){48}}
\put(72, 64){\tiny $ (m,n)$}
\put(144, 70){\tiny $ (m+2,n)$}
\put(76, 32){\tiny $ (m-1,n-1)$}
\put(76, 96){\tiny $ (m-1,n+1)$}
\put(72,0){\small{(b) Arrangement 2, $\ol Q_{m,n}$ in centre}, }
\put(87,-12){\small{neighbouring nodes are $\hat Q_{m+2,n}$, }}
\put(87,-24){\small{$\hat Q_{m-1,n+1}$, and $\hat Q_{m-1,n-1}$.}}
\end{picture}
\caption{Labelling of the nodes in the lattice. }
\label{fig:set of nodes}
\end{figure}

To derive governing equations of this lattice we introduce vectors 
${\bf e}_i=[2,0]^T = 2{\bf i}$, ${\bf e}_j=[-1,h]^T = h{\bf j} 
- {\bf i}$ and ${\bf e}_k=[1,h]^T = {\bf i} + h{\bf j}$, 
as shown on the right hand side of Figure \ref{fig:enlarged}, 
to describe the two configurations by which nodes are connected to 
nearest neighbours.
The honeycomb lattice is composed of two distinct arrangements of 
connecting nearest neighbour nodes, shown in Figure \ref{fig:set of nodes}.
We refer to Arrangement 1 as left-facing nodes, since they are connected 
to a nearest neighbour horizontally to the left. Arrangement 2 
will be referred to as right-facing nodes. When looking at Figure 
\ref{fig:set of nodes} we note that each node is connected to three 
nodes of the opposite arrangement.

\subsection{Derivation of the governing equations} 
\label{subsec:derivation} 

In the application we consider here, at every node there is a 
nonlinear capacitor, and between adjacent nodes, a linear inductor. 
We denote the voltage across the capacitor $(m,n)$ by $V_{m,n}$ 
and the total charge stored on this capacitor by $Q_{m,n}$.  
Finally, the current in the direction of the vector ${\bf e}_i$, 
through the inductor immediately to the  right of $(m,n)$ is 
denoted by $I_{m,n}$, and $J_{m,n}$ and $K_{m,n}$ represent 
currents in the direction of the vectors ${\bf e}_j$ and 
${\bf e}_k$, respectively. This configuration is illustrated in 
Figure \ref{fig:enlarged}, which shows an enlarged view 
of the lattice with the relevant currents indicated.

We derive separate governing equations for the two arrangements 
of nodes, and only later aim to reconcile the two into a single 
description. Our aim is to find equations for the variable, $Q_{m,n}$ 
at each node of the lattice.  To enable equations to be derived, we 
need to make the distinction between left- and right-facing nodes.  
We use $\hat{Q}_{m,n}$ for left-facing nodes, that is, arrangement 1 in 
Figure \ref{fig:set of nodes}(a), and $\bar{Q}_{m,n}$ for right-facing 
nodes, namely arrangement 2 in Figure \ref{fig:set of nodes}(b).  We use 
$Q_{m,n}$ for a general node, in practice, this will be either one of the 
left-facing ($\hat{Q}_{m,n}$) or the right-facing ($\bar{Q}_{m,n}$) nodes.  
A derivation from Kirchoff's laws has been given in \cite{Buttwatt2}. 
Here we simply quote the Hamiltonian 
\beqa 
\tilde{H} & = & \!\!\! \sum_{{{m,n ; s.t.}\atop{ \bar{P}_{m,n} {\rm exists}}} } \!\!\!
\half (\bar{P}_{m,n} \!-\! \hat{P}_{m\!+\!2,n})^2 + 
\half (\bar{P}_{m,n} \!-\! \hat{P}_{m\!-\!1,n\!+\!1})^2 + 
\half (\bar{P}_{m,n} \!-\! \hat{P}_{m\!-\!1,n\!-\!1})^2 \nn \\ && \qquad \qquad + 
\Upsilon(\bar{Q}_{m,n}) + \Upsilon(\hat{Q}_{m\!+\!2,n}),
\lbl{eqn:ham}  \eeqa
and note that $P_{m,n}$ and $Q_{m,n}$ are the conjugate momentum 
and displacement variables of the system.   The charge-voltage 
relationship is given by $V(Q_{m,n}) = \Upsilon'(Q_{m,n})$ where we 
assume the potential $\Upsilon(Q_{m,n})$ has the form $\Upsilon(Q) 
= \half Q^2 + \rec{3}a Q^3 + \rec{4}b Q^4$.  Since our analysis is based 
on small amplitude nonlinear expansions, we assume that there is a 
Taylor series of $\Upsilon(Q)$; we do not consider potentials of the 
form $\Upsilon(Q) \sim Q^\nu$ with $\nu<2$.   In Section 
\ref{subsec:estenergy} we use the Hamiltonian (\ref{eqn:ham}) to 
find the energy of small amplitude breathers. 

The lattice equations are obtained by eliminating $P_{m,n}$ 
from the equations 
\beq
\frac{d Q_{m,n}}{d t} = \frac{\partial H}{\partial P_{m,n}}, \qquad 
\frac{dP_{m,n}}{d t}=-\frac{\partial H}{\partial Q_{m,n}}=-\Upsilon'(Q_{m,n}) . 
\lbl{eqn:dhdq}  \eeq
Thus, for left-facing nodes we have 
\beqa 
\frac{d^{2}\hat{Q}_{m,n}}{d t^{2}} & = & \bar{Q}_{m-2,n} 
+ \bar{Q}_{m+1,n-1} + \bar{Q}_{m+1,n+1} - 3 \hat{Q}_{m,n} \nn \\ && 
+ a \bar{Q}_{m-2,n}^2 + a \bar{Q}_{m+1,n-1}^2 
+ a \bar{Q}_{m+1,n+1}^2  - 3 a \hat{Q}_{m,n}^2 \nn \\ && 
+ b \bar{Q}_{m-2,n}^3 + b \bar{Q}_{m+1,n-1}^3  
+ b \bar{Q}_{m+1,n+1}^3  - 3 b \hat{Q}_{m,n}^3 , 
\lbl{eqn:d2q1qonly}  \eeqa
where $m,n  \in {\rm Z\!\!Z}$, $\hat{Q}_{m,n}$ represents the 
charge at left-facing nodes and $\bar{Q}_{m,n}$ represents the 
charge at right-facing nodes. 
The right-facing nodes in arrangement 2 are governed by 
\beqa 
\frac{d^{2}\bar{Q}_{m,n}}{d t^{2}} & =  & \hat{Q}_{m+2,n} 
+ \hat{Q}_{m-1,n+1} + \hat{Q}_{m-1,n-1} - 3 \bar{Q}_{m,n} \nn \\ && 
+ a \hat{Q}_{m+2,n}^2 + a \hat{Q}_{m-1,n+1}^2 
+ a \hat{Q}_{m-1,n-1}^2 -3 a \bar{Q}_{m,n}^2 \nn \\ && 
+ b \hat{Q}_{m+2,n}^3 + b \hat{Q}_{m-1,n+1}^3 
+ b \hat{Q}_{m-1,n-1}^3 -3 b \bar{Q}_{m,n}^3 .
\lbl{eqn:d2q2qonly}  \eeqa

\section{General theory}
\label{sec:gen}

\subsection{Asymptotic analysis}
\label{subsec:asymp}

We now aim to find an approximate analytic solution to the equations 
(\ref{eqn:d2q1qonly}) and (\ref{eqn:d2q2qonly}) by applying the method 
of multiple scales. We first rescale the current variables $m,n$ and $t$,  
introducing the new variables
\beq
X = \varepsilon m, \qquad  Y = \varepsilon hn, \qquad  
\tau = \varepsilon t , \qquad {\rm and}  \qquad   T = \varepsilon^2 t , 
\lbl{eqn:variables} \eeq
with $\varepsilon \ll 1$ being the amplitude of the breather, 
the variables $X,Y$ will be treated as continuous real variables. 

We require different ansatzes for the right- and left-facing nodes, 
therefore we analyse each type of node individually using 
(\ref{eqn:d2q1qonly}) and (\ref{eqn:d2q2qonly}). 
For right-facing nodes we seek solutions of the form
\beqa
\bar{Q}_{m,n}(t) & = &  \ {\varepsilon} \ee^{i\psi} F(X,Y,\tau, T) 
+ {\varepsilon}^2 \left[ G_{0}(X,Y,\tau, T) +  \ee^{i\psi} G_{1}(X,Y,\tau, T) 
\right. \nn \\ && \left. + \ee^{2i\psi} G_{2}(X,Y,\tau, T) \right] 
+ {\varepsilon}^3 \left[ H_{0}(X,Y,\tau, T) + \ee^{i\psi} H_{1}(X,Y,\tau, T) 
\right. \nn \\ && \left. 
+  \ee^{2i\psi} H_{2}(X,Y,\tau, T) + e^{3i\psi} H_{3}(X,Y,\tau, T) \right] 
+ ... + c.c.,
\lbl{eqn:expanofqmnright} \eeqa
where the phase of the carrier wave $\psi$ is given by $\psi = 
km + lhn + \omega t$, where ${\bf k}=[k,l]^T$ is the wavevector 
and $\omega({\bf k})$ is its temporal frequency. Similarly, for 
left-facing nodes we seek solutions of the form
\beqa
\hat{Q}_{m,n}(t) & = &  {\varepsilon}e^{i\psi} P + {\varepsilon}^2 \left[ 
Q_{0} + \ee^{i\psi} Q_{1} + \ee^{2i\psi} Q_{2} \right] \nn \\ && 
+ {\varepsilon}^3 \left[ R_{0} + \ee^{i\psi} R_{1} + \ee^{2i\psi} R_{2} 
+ \ee^{3i\psi} R_{3}  \right] + ... + c.c., 
\lbl{eqn:expanofqmnleft} \eeqa 
where $P,Q_j,R_j$ are all functions of $(X,Y,\tau,T)$. 

After substituting the ansatzes (\ref{eqn:expanofqmnright}) and 
(\ref{eqn:expanofqmnleft}), into the relevant right- and left-facing 
lattice equations (\ref{eqn:d2q1qonly}) and (\ref{eqn:d2q2qonly}), 
we equate the coefficients of each harmonic frequency at each 
order of $\varepsilon$ to find two sets of equations, which we 
analyse in order below.   We use the slightly unusual notation 
${\cal O}(\ep^p \ee^{iq\psi})$ to mean those terms of ${\cal O}(\ep^p)$ 
which have the coefficient $\ee^{iq\psi}$, that is, we neglect those 
terms which have $\ee^{i r \psi}$ with $r\neq q$.  

Since our main calculations are only going as far as ${\cal O}(\ep^3)$ and 
${\cal O}(\ep^4\ee^{0i\psi})$, the variables (\ref{eqn:variables}) are 
sufficient for our analysis.   At higher orders of $\ep$, we would 
have to include longer space scales, given by $\tilde X = \ep^2 m$ 
and $\tilde Y = \ep^2 h n$, however, these make little difference to 
the shape of the breather, as shown in \cite{jadw-hot}.  

\subsection{${\cal O}(\ep\ee^{i\psi})$ - the dispersion relation}
\label{subsec:disp}

The first order we investigate is $\cal{O}(\ep \ee^{i \psi})$, 
whence we obtain 
\beq 
{\bf M} \left( \begin{array}{c}  F \\  P  \end{array} \right) = 
\left( \begin{array}{cc} 3 - \omega^2  & -\beta \\ 
-\beta^* & 3 - \omega^2 \end{array}  \right) \left( 
\begin{array}{c}  F \\  P  \end{array} \right) = \bf{0} , 
\lbl{mat:fp} \eeq
where 
\beq
\beta= e^{2ik} + e^{-ik-ilh} +e^{-ik+ilh} , 
\lbl{beta-def} \eeq
and $\beta^*$ is its complex conjugate. We write $\beta = |\beta| 
\ee^{-i\theta}$, the magnitude being
\beq
|\beta| = \sqrt{3+ 2\cos (2lh) + 2\cos(3k+lh) + 2\cos (3k-lh)} . 
\lbl{eqn:mb} \eeq

We are interested in solutions where $( {{F}\atop{P}}) \neq \bf{0}$, 
equation (\ref{mat:fp}) is thus an eigenvalue problem. We require the 
determinant of the matrix to be zero, which gives the dispersion relation
\beq
\omega^2 = 3 \pm \sqrt{3+2\cos (2lh)+2\cos(3k+lh)+2\cos (3k-lh)} . 
\lbl{eqn:dispersionrelation} \eeq
The dispersion relation describes the dependence of the temporal 
frequency of the wave on the wavenumbers $(k,l)$. 

The negative square root in (\ref{eqn:dispersionrelation}) leads 
to an `acoustic' branch, or surface in $(k,l,\omega)$ space with 
lower frequencies, which we denote by $\omega_{ac}=\sqrt{3-|\beta|}$; 
and we have $\omega_{ac} \rightarrow 0$ as $k,l \rightarrow 0$. 
The surface corresponding to the positive root in 
(\ref{eqn:dispersionrelation}), which clearly has larger values of 
$\omega$, we denote by $\omega_{opt}=\sqrt{3+|\beta|}$, and we 
describe this surface as the `optical' branch. The acoustic branch 
accounts for frequencies in the range $0 \leq \omega \leq \sqrt{3}$, 
whilst the optical branch satisfies $\sqrt{3} \leq \omega \leq \sqrt{6}$. 

\begin{figure} [ht] 
\includegraphics[width=12cm]{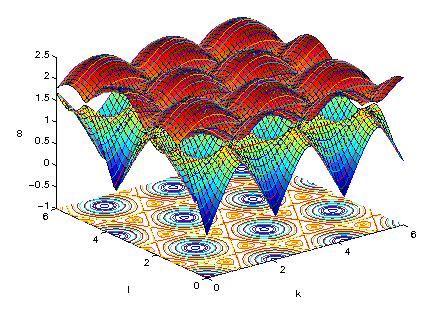}
\caption{Plot of $\omega$(\bf{k}), (in colour in online version).}
\label{fig:plotw}
\end{figure}

The plot of $\omega$ against $k$ and $l$  along with the 
contour plot is shown in Figure \ref{fig:plotw}. 
We have the dispersion relation (\ref{eqn:dispersionrelation}) 
for the two coupled systems (\ref{eqn:d2q1qonly}) and 
(\ref{eqn:d2q2qonly}). We consider $k$ and $l$ such that 
$(k,l) \in T^2 = [0,2\pi] \times [0,2\pi/h]$  because $\omega$ 
is periodic in both $k$ and $l$, with period $2\pi$ in the 
$k$-direction and period $2\pi/h$ in $l$-direction.

\begin{figure} [ht] 
{ \includegraphics[width=4cm]{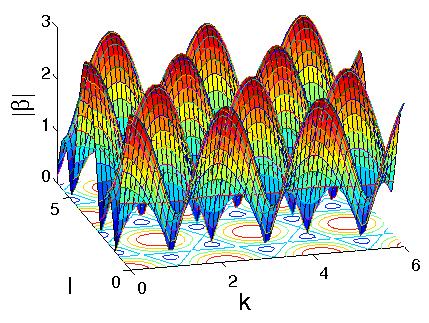} \ 
\includegraphics[width=4cm]{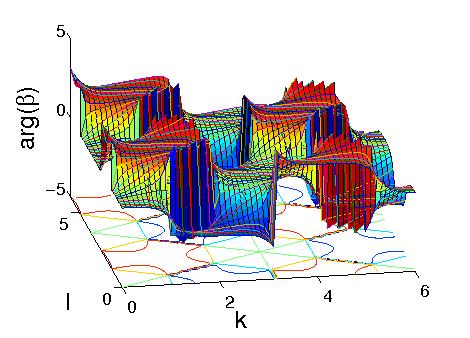} \ 
\includegraphics[width=4cm]{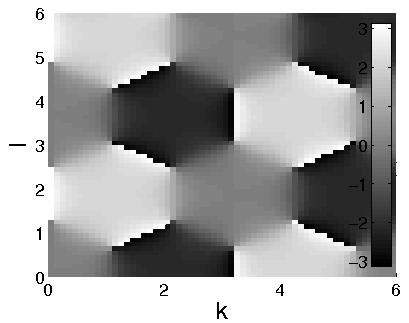}}
\caption{Left: plot of $|\beta|$; centre: plot of arg$(\beta)$, 
(both in colour in online version); right: greyscale plot of arg$(\beta)$, 
showing large regions where arg$(\beta) \approx 0, \pm 2\pi/3$. }
\label{beta-fig}
\end{figure}

The locations in $(k,l)$ space of the minima of $\omega_{ac}$ and 
the maxima of $\omega_{opt}$ coincide are seen in Figure 
\ref{fig:plotw} as the circles in the centres of the hexagonal shapes 
in the contour plot.   These points are at $(0,0)$, $(2\pi/3,0)$, 
$(0,2\pi/h)$, $(2\pi/3,2\pi/h)$ and $(\pi/3,\pi/h)$ etc.   Points where 
the two surfaces meet are also evident in Figure \ref{fig:plotw} as 
the centres of the triangles surrounding the hexagonal shapes, 
at these points where  $\omega = \sqrt{3}$.  The $\omega(k,l)$ 
dispersion surfaces have cusp-like singularities at these points,  
which can be denoted by $\bf{k}_1$,...,$\bf{k}_6$, where
\beq \begin{array}{rclcrcl}
\textbf{k}_1 &=& [\pi/3,\pi/3h ]^T, && 
\textbf{k}_2 &=& [\pi, \pi/3h]^T ,\\
\textbf{k}_3 &=& [0, 2\pi/3h]^T , && 
\textbf{k}_4 &=& [0, 4\pi/3h]^T,\\
\textbf{k}_5 &=& [2\pi/3, 2\pi/3h]^T, && 
\textbf{k}_6 &=& [2\pi/3, 4\pi/3h]^T.
\end{array} \lbl{eqn:wavevectors} \eeq 
By comparing (\ref{eqn:dispersionrelation}) with (\ref{eqn:mb}), 
we observe that these points occur where $\beta=0$. 
In graphene, 
these wavevectors are known as Dirac points \cite{peleg}.  
Figure \ref{fig:plotw} also illustrates the hexagonal symmetry 
of the lattice.   Figure \ref{beta-fig} shows the magnitude and 
argument of $\beta$ as function of $(k,l)$.  Note the presence 
of sizable plateaus where arg$(\beta) \approx 0, \pm 2\pi/3$. 

However, equation (\ref{mat:fp}) remains unsolved.  
Since det$({\bf M})$=0, solutions can be written as $P = CF$, 
where, for $\omega_{ac}$ and $\omega_{opt}$, we have 
\beq 
C_{ac}    
= \frac{\beta^*}{|\beta|}   
= \ee^{i\theta} , 
\qquad C_{opt} = -C_{ac} = -\ee^{i\theta} , 
\lbl{eqn:C} \eeq
respectively,  the latter expression arising from $\beta = 
\Gamma \ee^{-i\theta}$.   These expressions for $C_{ac}$, 
$C_{opt}$ will be used in later calculations, where we find 
expressions for the functions $G_2$, $G_1$, $G_0$, 
$Q_2$, $Q_1$ and $Q_0,$ in terms of $F$.

\subsection{${\cal O}(\ep^2\ee^{0i\psi})$: relationship between 
$G_0$ and $Q_0$}
\label{subsec:ep2e0}

At ${\cal O}(\ep^2 \ee^{0i\psi})$, we obtain the same equation 
from both (\ref{eqn:d2q1qonly}) and (\ref{eqn:d2q2qonly}), 
which are the equations for $\hat Q_{m,n}$ and for $\ol Q_{m,n}$. 
\beq
G_0 + G_0^* + 2a|F|^2 = Q_0 + Q_0^* + 2a|P|^2 .  
\lbl{ep2ex0eq} \eeq 
Note that from the ansatz, $Im(G_0)$ is irrelevant, since only 
the combination $G_0+G_0^*$ ever appears in our equations.  
Hence, we assume $Im(G_0)=0$, and only consider the real 
parts, that is, $G_0=G_0^*$.    Since $P=CF$ with $|C|=1$, 
we have $|F|^2=|P|^2$ in (\ref{ep2ex0eq}), 
and $G_0=Q_0$, but this quantity is not yet determined. 

\subsection{${\cal O}(\ep^2\ee^{2i\psi})$: expressions for $G_2$ and $Q_2$}
\label{subsec:e2epsilon2}

As previously mentioned, we aim to express all the variables 
$G_0, G_1, G_2, Q_0, Q_1$ and $Q_2$ in terms of $F$. 
At  $\cal{O}$ $(\varepsilon^2 e^{2i \psi})$ by substituting the ansatzes 
(\ref{eqn:expanofqmnright})--(\ref{eqn:expanofqmnleft}) into
(\ref{eqn:d2q1qonly})--(\ref{eqn:d2q2qonly}), we obtain 
\beqa
(3-4\omega^2) G_2 - \gamma Q_2 & = & \gamma a P^2 - 3 a F^2 ,
\lbl{eqn:omega4G2} \\
(3-4\omega^2) Q_2 -\gamma^* G_2&=&\gamma^* aF^2- 3a P^2 ,
\lbl{eqn:omega4Q2} \eeqa 
where $\gamma^*$ is the complex conjugate of $\gamma$ and
\beq
\gamma = e^{4ik} + e^{-2ik+2ilh} + e^{-2ik-2ilh}  . 
\lbl{eqn:t1}  \eeq 
Note that if we think of $\beta$ and $\gamma$ as being functions of 
$(k,l)$, they are related by $\gamma(k,l) = \beta(2k,2l)$, compare 
(\ref{beta-def}) and (\ref{eqn:t1}). 
Solving the linear system (\ref{eqn:omega4G2})--(\ref{eqn:omega4Q2}) 
for $G_2$ and $Q_2$ as functions of $F$, we find 
\beq
\left( \begin{array}{c}  G_2 \\  Q_2  \end{array} \right) =
\frac{a C F^2}{ (3 - 4\omega^2)^2 - |\gamma|^2}  
\left( \begin{array}{c}
(|\gamma|^2 -9 + 12 \omega^2) C^* - 4 \omega^2 \gamma C \\ 
(|\gamma|^2 -9 + 12 \omega^2) C - 4 \omega^2 \gamma^* C^* 
\end{array} \right) .  \lbl{mat:g2q2} \eeq
Whilst the bottom term in the vector is the complex conjugate of the top, 
we do not have $Q_2=G_2^*$ since the the term $CF^2$ common to 
both is not necessarily real. We return to the expressions 
(\ref{mat:g2q2}) in Section \ref{subsec:asym}.

\subsection{${\cal O}(\ep^2\ee^{i\psi})$: the velocity profile}
\label{subsec:velocity}

We now consider the governing equations at ${\cal O}(\ep^2 \ee^{i\psi})$, 
which can be written as 
\beq
\textbf{M} \left( \begin{array}{c} G_1 \\ Q_1 \end{array} \right) = 
\left( \begin{array}{cc} -2i{\omega}F_{\tau} - i \beta_k P_X -i \beta_l P_Y \\
-2i{\omega}P_{\tau} - i\beta_k^*F_X - i\beta_l^*F_Y \end{array} \right) ,
\lbl{mat:e2eps1} \eeq
where $\textbf{M}$ is the matrix given in (\ref{mat:fp}), 
and $\beta_k$, $\beta_l$ are the partial derivatives of $\beta$ 
with respect to $k$, $l$ respectively, namely 
\beq
\beta_k = 2 i \ee^{2ik} - i \ee^{-ik-ilh} - i \ee^{-ik+ilh} , \qquad 
\beta_l = i h \ee^{-ik+ilh} - i h \ee^{-ik-ilh} . 
\lbl{betak-betal} \eeq 

Since det$({\bf M})=0$, an equation such as (\ref{mat:e2eps1}), 
which we write as ${\bf M}  ({{G_1}\atop{Q_1}}) = {\bf d}$, 
either has no solutions, or a whole family of solutions for $(G_1,Q_1)^T$. 
According to the Fredholm alternative, the existence of solutions 
depends on ${\bf d}$.  Solutions exist only if the {\sc rhs} of 
(\ref{mat:e2eps1}), namely ${\bf d}$, is in the range of the matrix 
${\bf M}$, which is given by 
\beqa
{\rm Range}_{ac}
&=& K \left( \begin{array}{c} -\beta \\ |\beta| \end{array} \right)  
= K \left( \begin{array}{c} -\ee^{-i\theta} \\ 1 \end{array} \right) , 
\nn \\ {\rm Range}_{opt} 
&=& K \left( \begin{array}{c} \beta \\ |\beta| \end{array}  \right)  
= K \left( \begin{array}{c} \ee^{-i\theta} \\ 1 \end{array} \right) . 
\eeqa
Since normals to these directions are given by 
\beq 
{\bf n}_{ac} = \left( \begin{array}{c} \ee^{i\theta} \\ 1 \end{array} \right) , \qquad 
{\bf n}_{opt}=\left( \begin{array}{c} -\ee^{i\theta} \\ 1 \end{array} \right) , 
\lbl{ndef} \eeq 
the condition that ${\bf d}\in$Range implies ${\bf n.d}=0$. 
Note that in both the optical and the acoustic cases, (\ref{ndef}) 
implies ${\bf n} = ({{C}\atop{1}})$. 

\begin{figure}[ht] 
 {\centering  
\includegraphics[width=80mm]{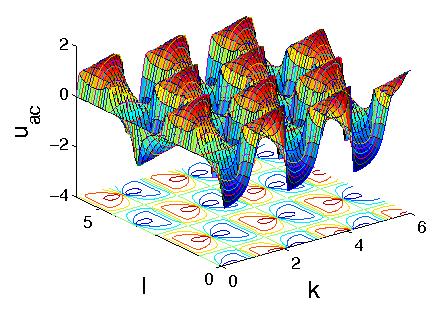} \ 
\includegraphics[width=80mm]{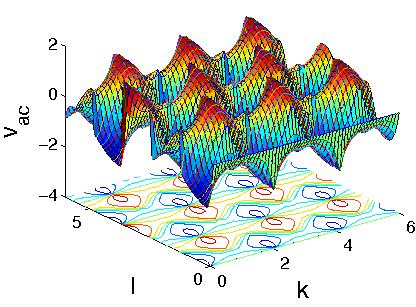}}
\\
{\centering  
\includegraphics[width=80mm]{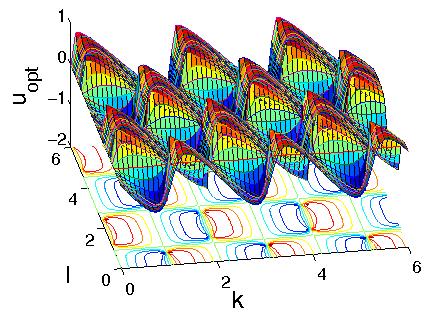} \ 
\includegraphics[width=80mm]{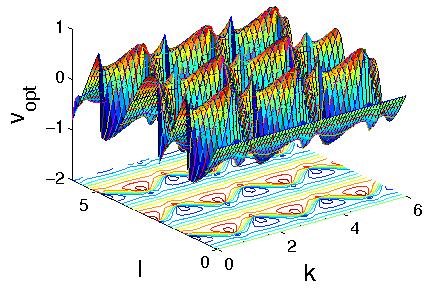}}
\caption{Plot of the horizontal ($u$, on the left) and vertical 
($v$, on the right) components of velocity, both as functions 
of the wavenumbers $k$, $l$; upper row, acoustic mode; 
lower row, optical mode, (in colour in online version). }
\label{fig:uv}
\end{figure}

We also recall that the leading order quantities, $P$ and $F$ are related 
by $P=CF$, where both $C$ and ${\bf n}=({{C}\atop{1}})$ 
have different expressions for the acoustic and optical cases, given by 
(\ref{eqn:C}).  Using $P=CF$ and ${\bf n.d}=0$ we obtain the equation 
\beq
0 = 4\omega F_\tau + (\beta_k C + \beta_k^* C^*) F_X 
+ ( \beta_l C + \beta_l^* C^* ) F_Y .
\lbl{eqn:velocity2}
\eeq
This equation implies that $F$ (and hence $P$ as well) is a travelling 
wave.  We write 
\beq
F(X,Y,\tau,T)\equiv F(Z,W,T), \quad {\rm where} \;\;\;  
Z = X - u\tau, \quad  W = Y - v\tau , 
\lbl{tw} \eeq
the horizontal and vertical velocity components are found to be
\beqa 
u & = & \frac{\beta_k C + \beta_k^* C^*}{4\omega} 
= \frac{-3\sin(3k)\cos(lh) }{\omega|\beta|}, \lbl{eqn:u} \\ 
v & = & \frac{\beta_l C + \beta_l^* C^*}{4\omega} 
= \frac{-h\sin(lh) (\cos(3k)+2\cos(lh))}{\omega|\beta|}.
\lbl{eqn:v} \eeqa 
As expected from the standard theory of waves \cite{whitham}
these are simply the derivatives of the frequency with respect to the 
wavenumber, $u=\partial\omega/\partial k$, $v=\partial\omega/\partial l$. 
Since we have different expressions for $\omega_{ac}$ and 
$\omega_{opt}$, equations (\ref{eqn:u})--(\ref{eqn:v}) also 
generates different 
formulae for $u_{ac}$ and $u_{opt}$ (and $v_{ac}$ and $v_{opt}$). 
Figure \ref{fig:uv} shows plots of the horizontal and vertical 
components of the velocity as functions of the wavenumbers $(k,l)$.  
Note that at the Dirac points, where $\beta=0$, the singularity 
is removable since the numerators in (\ref{eqn:u})--(\ref{eqn:v}) 
are also zero. 

\begin{figure}[ht] 
 {\centering  
 \includegraphics[width=80mm]{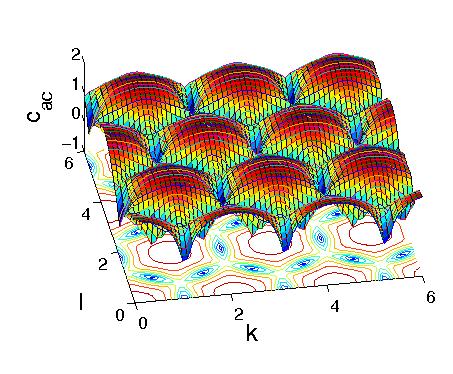} \ 
 \includegraphics[width=80mm]{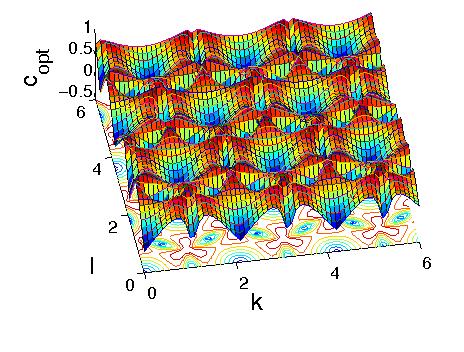} }
\caption{Plot of the total speed as a function of the wavenumbers $(k,l)$; 
left, acoustic case, right, optical mode, (in colour in online version). }
\label{fig:c}
\end{figure}

The overall speed, $c$, is given by   
\beqa 
c &=& \sqrt{u^2 + v^2}  = \frac{h}{\omega|\beta|} 
\sqrt{ \sin^2(lh) [\cos(3k)+2\cos(lh)]^2 + 3\sin^2(3k)\cos^2(lh) } , 
\nn \\ && \lbl{eqn:avevelocity} 
\eeqa 
which is plotted in figure \ref{fig:c}.  Whilst the above calculations, 
(\ref{eqn:u})--(\ref{eqn:avevelocity}), are for the acoustic mode, 
similar calculations for the optical mode produce similar results. 
All these plots show periodic behaviour, however, the hexagonal 
symmetry of the system only becomes clear in the total speed,  
the plots of the velocities $u,v$ have a more complicated, 
although complimentary form.  The velocities $u,v$ both show 
sensitive dependence on wave vector $(k,l)$.  At the wavevectors 
$\textbf{k}_1$,...,$\textbf{k}_2$, found in (\ref{eqn:wavevectors}), 
both the components of velocity are zero.

The above calculation gives the condition on the {\sc rhs} of 
(\ref{mat:e2eps1}) for solutions to exist.  However, the quantities 
$G_1$, $Q_1$ remain unknown. 
The solutions of (\ref{mat:e2eps1}) are degenerate, and the 
one-parameter family of solutions may be written as 
\beq
\left( \begin{array}{c} G_1 // Q_1 \end{array} \right) = 
\left( \begin{array}{c} \ol G_1 + \hat G_1 // C\ol G_1 
\end{array} \right)  , \lbl{G1ansatz} 
\eeq 
for arbitrary $\ol G_1$.  The two equations for $\hat G_1$ 
from (\ref{mat:e2eps1}) are then identical, and are solved by 
\beqa
\hat G_1 & = & \frac{- i s }{2|\beta|} \left[  (\beta_k C - \beta_k^* C^*) F_Z 
+ (\beta_l C - \beta_l^* C^*) F_W \right] , \quad 
s = \left\{ \begin{array}{lll} +1 && {\rm acoustic, } \\ 
-1 && {\rm optical. } \end{array} \right. \nn \\ && 
\lbl{hatG1} \eeqa 
Writing this as $\hat G_1 = \hat u F_Z + \hat v F_W$, we have 
\beq
\hat u = \frac{2}{|\beta|^2} \left[ \cos (3k) \cos(lh) - \cos(2lh)
\right] , \qquad  \hat v = \frac{2 h}{|\beta|^2} \sin(3k)\sin(lh) . 
\lbl{eqn:uhatvhat} \eeq
Whilst, this leaves $\ol G_1$ undetermined, the quantity 
$\hat G_1$ describes a small difference in the evolution of 
the left- and right-handed nodes of the honeycomb lattice.  

\subsection{${\cal O}(\ep^3\ee^{0i\psi})$: corrections to the slow mode}
\label{subsec:HoRo}

At ${\cal O}(\ep^3 \ee^{0i\psi})$, we obtain the equation 
\beq
0 = 3 H_0 - 3 R_0 + 3 a (F^*G_1 + FG_1^*) - 3 a (P^*Q_1+PQ_1^*) , 
\eeq
from the substitution of
(\ref{eqn:expanofqmnright})--(\ref{eqn:expanofqmnleft}) into both 
(\ref{eqn:d2q1qonly}) and (\ref{eqn:d2q1qonly}).   
We are only interested in determining the leading order terms 
$\ep F$, $\ep P$, which require the ${\cal O}(\ep^2)$ terms 
$G_0,Q_0$, so we do not pursue the determination of $H_0,R_0$, 
which are ${\cal O}(\ep^3)$ correction terms and provide only 
a small difference between the right-facing and left-facing nodes. 

\subsection{${\cal O}(\ep^4\ee^{0i\psi})$: expressions for $G_0$ and $Q_0$}
\label{subsec:Go Qo}

In determining $G_0$ and $Q_0$, in Section \ref{subsec:ep2e0} we 
found a single relationship at $\cal{O}$ $(\varepsilon^2 e^{0})$, namely 
$G_0=Q_0$. At ${\cal O}(\varepsilon^3 e^{0})$, we again obtained a 
single equation; we now move on to consider ${\cal O}(\varepsilon^4 e^{0})$. 
%
%
Noting that $G_0=Q_0$ and $G_0^*=G_0$, $|F|^2=|P|^2$, 
and consequent results, such as $|G_2|^2=|Q_2|^2$, 
allows significant simplification. 
Furthermore, the ${\cal O}(\ep^3\ee^{i\psi})$ 
equations analysed below in section \ref{subsec:nls}, 
have the same form as those in Section \ref{subsec:velocity}, 
namely, $H_1,R_1$ satisfy a system of the form ${\bf M} 
( {{H_1}\atop{R_1}} ) = ( {{A}\atop{B}} )$ 
for some $A,B$ where ${\bf M}$ is singular. Hence we write 
the solution for $H_1,R_1$ as $({{H_1}\atop{R_1}} ) =
({{\ol H}\atop{C\ol H}} )  + ( {{\hat H}\atop{0}} ) $. 
Ultimately we obtain the equation for $G_0$ as 
\beq
G_{0\tau\tau} = 3 \nabla^2 ( G_0 + a |F|^2 ) - 3 a ( |\hat G|^2 + 
\hat G^* \ol{G} + \hat G \ol{G}^* + F^* \hat H + F \hat H^* ) . 
\lbl{goeq} \eeq 

In general, we cannot solve (\ref{goeq}) to find $G_0$ and 
$Q_0$, but there are two special cases when we can do so. 
In the cases considered later, either $\hat H=0$ or it is not 
relevant to our calculations. We also choose $\ol G =-\half \hat G$ 
with (\ref{hatG1}), so that (\ref{goeq}) can be simplified to 
$G_{0,\tau\tau} =3 \nabla^2 (G_0 + a |F|^2)$, 
which is similar to the previously derived results for the 
square and hexagonal lattices, see equations (2.23) of 
\cite{Buttwatt1} and (2.23) of \cite{Buttwatt2}. 

The first special case, analysed in Section \ref{subsec:sym}, is when 
the interaction potential is symmetric, that is, $V(Q_{m,n})=V(-Q_{m,n})$.  
Under this assumption, the quadratic coefficient of the force, $a$, is zero, 
leading to $G_0=Q_0=0$ as the solution of (\ref{goeq}). In this case, we 
also have $G_2=Q_2=0$.  This means that the system is governed by 
equations (\ref{mat:e3eps1})--(\ref{Bdef}) given below, which can be 
reduced to a single NLS equation. 

In Section \ref{subsec:asym}, we consider the second case, $\omega = 
\omega _{max}$ where $\omega_{max}$ represents the maxima of $\omega$.  
In this case the breather is stationary, since (\ref{eqn:u})--(\ref{eqn:v}) 
yield $u=0$ and $v=0$, then the system as a whole has no 
$\tau$-dependence, that is, $P_{\tau\tau} = F_{\tau\tau}=0$.  
In this case, $G_0,Q_0$ also have no $\tau$-dependence, and 
so equation (\ref{goeq}) reduces to $G_0=Q_0=-a|F|^2$.   This solution 
can be substituted into equations (\ref{mat:e3eps1})--(\ref{Bdef}) 
given below which again can be reduced to a single NLS equation.  

\subsection{Nonlinear Schr\"{o}dinger equation}
\label{subsec:nls}

The final equation we need to investigate comes from terms of 
${\cal O}(\varepsilon^3 \ee^{i \psi})$ which yield 
\beq 
{\bf M} \left( \begin{array}{c}  H_1 \\  R_1  \end{array} \right)
 = \left( \begin{array}{c}  A  \\  B  \end{array} \right),
\lbl{mat:e3eps1} \eeq
where the matrix ${\bf M}$ is identical to that in (\ref{mat:fp}), 
and the {\sc rhs} components are given by 
\beqa
A & = & -2 i w G_{1\tau} - i \beta_k Q_{1,X} - i \beta_l Q_{1,Y} 
- 2 i w F_T - F_{\tau\tau} \nn \\ && 
+ \half P_{XX} (4\ee^{2ik}+\ee^{-ik+ilh}+\ee^{-ik-ilh}) 
+ h P_{XY} ( \ee^{-ik-ilh} - \ee^{-ik+ilh} ) \nn \\ && 
+ \mfrac{3}{2} P_{YY} ( \ee^{-ik+ilh}+\ee^{-ik-ilh} ) 
+ 3 \beta b |P|^2 P - 9 b |F|^2 F \nn \\ && 
+ 2 a [ \beta (PQ_0+PQ_0^*+P^*Q_2) - 3(FG_0+FG_0^*+F^*G_2) ] ,
\lbl{Adef} \\
B & = & -2 i w Q_{1\tau} - i \beta^*_k G_{1,X} - i \beta^*_l G_{1,Y} 
- 2 i w P_T - P_{\tau\tau} \nn \\ && 
+ \half F_{XX} ( 4\ee^{-2ik}+ \ee^{ik-ilh} + \ee^{ik+ilh} )  
+ h F_{XY} ( \ee^{ik+ilh} - \ee^{ik-ilh} ) \nn \\ &&
+ \mfrac{3}{2} F_{YY} ( \ee^{ik-ilh} + \ee^{ik+ilh} ) 
+ 3 \beta^* b |F|^2 F - 9 b |P|^2 P \nn \\ && 
+ 2 a [ \beta (FG_0+FG_0^*+F^*G_2) - 3(PQ_0+PQ_0^*+P^*Q_2)  ] . 
\lbl{Bdef} \eeqa

As in the case of the equations at ${\cal O}(\ep^2 \ee^{i\psi})$, 
in order for this system of equations to have solutions, there 
is the consistency condition ${\bf n}.({{A}\atop{B}})=0$ 
that must be satisfied.  

An equation for $F(Z,W,T)$ can be obtained from
(\ref{mat:e3eps1})--(\ref{Bdef}) by the following procedure:  
\\ \phantom{.} \qquad (i) calculating the consistency condition 
                                   ${\bf n}.({{A}\atop{B}})=0$,  
				   that is, $CA+B=0$, 
\\ \phantom{.} \qquad (ii) substitute in expressions for $G_2$, 
$Q_2$, $G_0$, $Q_0$, 
 \\ \phantom{.} \qquad (iii) making the substitution $P=CF$, 
\\ \phantom{.} \qquad (iv) transforming to travelling wave 
coordinates by (\ref{tw}).  \\ 
However, in general, this equation will still be coupled to 
$G_0$, through (\ref{goeq}); and carrying out this 
procedure in general leads to extremely lengthy expressions. 

\subsection{Summary} 
\label{subsec:summary}

We have derived a multiple scales asymptotic expansion for 
envelope solutions of the scalar two-dimensional honeycomb lattice. 
After finding the usual expressions for the frequency, and group velocity, 
we have found a coupled system of PDEs for the shape of the envelope 
given by (\ref{goeq}) and (\ref{mat:e3eps1})--(\ref{Bdef}).   Whilst we 
cannot, in general,  solve this resulting system of equations, there are 
two special cases in which the system reduces to a single NLS equation.  
The general case shares some similarities with the Davey-Stewartson 
system of equations \cite{DS} obtained in fluid mechanics. 
The remainder of this paper is {\em not} directed to a general analysis 
of equations (\ref{goeq}) and (\ref{mat:e3eps1})--(\ref{Bdef}), 
rather we consider two special cases in more detail. 

These two cases are considered in more detail in Sections 
\ref{subsec:sym} and \ref{subsec:asym} respectively, and in both cases 
the procedure (i)--(iv) leads to substantially simpler expressions than the 
general case.  In both special cases we find additional criteria which, if 
not satisfied, mean the the lattice cannot support breather solutions. 

\section{The symmetric potential ($a=0$) and moving breathers} 
\label{subsec:sym}

In this Section we consider the simplified case where $a = 0$ 
in (\ref{eqn:d2q1qonly}) and (\ref{eqn:d2q2qonly}) so that 
$V(-\phi)=V(\phi)$ and $V'(-\phi)=-V'(\phi)$. In section  
\ref{subsec:e2epsilon2}, whilst equation (\ref{mat:g2q2}) 
remains valid, we recover $G_2=Q_2=0$, that is, there is no 
generation of second harmonics. Furthermore, from Section 
\ref{subsec:Go Qo} we gain $G_0=0=Q_0$, which satisfies 
the relationship $G_0=Q_0$ from Section \ref{subsec:ep2e0}. 
This means that there is no `slow' mode, which is independent 
of $t$ (corresponding to $\omega=0$), and the localised mode 
evolves only on the slower $\tau,T$ timescales.   

\subsection{${\cal O}(\ep^3 \ee^{i\psi})$ - derivation of NLS}
\label{subse:sym-nls}

We now apply the procedure (i)--(iv) from Section \ref{subsec:nls} 
and so simplify the NLS-like system (\ref{mat:e2eps1})--(\ref{Bdef}). 
Taking $G_1$, $Q_1$ as given by (\ref{G1ansatz}) with 
$\ol G=-\half \hat G$ and using $P=CF$ to  evaluate 
${\bf n}.({{A}\atop{B}})= 0$, we obtain a single 
equation for $F$, namely 
\beqa
0 & = & - 4 i w F_T - 2 F_{\tau\tau} 
+ \half F_{XX} \left[ C (4\ee^{2ik}+\ee^{-ik+ilh}+\ee^{-ik-ilh}) 
\right. \nn \\ && \left. 
+ C^*( 4\ee^{-2ik}+ \ee^{ik-ilh} + \ee^{ik+ilh} ) \right]  \nn \\ && 
+ h F_{XY} \left[ C( \ee^{-ik-ilh} - \ee^{-ik+ilh} ) 
+ C^* ( \ee^{ik+ilh} - \ee^{ik-ilh} ) \right]  \nn \\ && 
+ \mfrac{3}{2} F_{YY} \left[ C( \ee^{-ik+ilh}+\ee^{-ik-ilh} ) 
+ C^* ( \ee^{ik-ilh} + \ee^{ik+ilh} ) \right] 
\nn \\ && 
+ 3 b |F|^2 F (\beta C+\beta^*C^* - 6) .  
\lbl{eqn:synls} \eeqa
Thus we have completed stages (i)--(iii) of the procedure 
from Section \ref{subsec:nls}. 
In stage (iv) we eliminate the $\tau$-derivative terms 
using the travelling wave substitution $F(X,Y,\tau,T) = 
F(Z,W,T)$ with $u$ and $v$ representing the horizontal 
and vertical components of velocity found in 
(\ref{eqn:u})--(\ref{eqn:v}). Using  $F_{\tau\tau} = u^2 F_{ZZ} 
+ 2 u v F_{ZW} + v^2 F_{WW}$, we rewrite (\ref{eqn:synls}) as
\beqa
4 i w F_T & = & D_Z F_{ZZ} + D_W F_{WW} + D_M F_{WZ} 
+ 3 b |F|^2 F (\beta C+\beta^*C^* - 6) , \lbl{nls:WZ}
\eeqa
where
\beqa
\Delta_Z &\!=\!& \frac{2}{|\beta|} \left[ 4 \cos(2lh)+5\cos(lh)\cos(3k)\right] 
, \qquad \Delta_M = \frac{4h}{|\beta|} \sin(3k)\sin(lh) , \nn \\ 
\Delta_W &\!=\!& \frac{6}{|\beta|} \cos(lh) \left[\cos(3k)+2\cos(lh)\right] , 
\lbl{Delta-def}\eeqa
\beqa \begin{array}{rclcrcl} 
D_{Z,ac} & = & \Delta_Z - 2 u_{ac}^2 - |\beta| \hat u^2 , && 
D_{Z,opt} & = & - \Delta_Z - 2 u_{opt}^2 + |\beta| \hat u^2 , \\ 
D_{W,ac} & = & \Delta_W - 2 v_{ac}^2 - |\beta| \hat v^2 , &&
D_{W,opt} & = & - \Delta_W - 2 v_{opt}^2 + |\beta| \hat v^2 , \\
D_{M,ac} & = & - \Delta_M - 4 u_{ac} v_{ac} - |\beta| \hat u^2 , && 
D_{M,opt} & = & \Delta_M - 4 u_{opt} v_{opt} + 2|\beta| \hat u \hat v . 
\end{array} && \nn \\ && \lbl{Dacdef} \eeqa
Hence, from the governing equations 
(\ref{eqn:d2q1qonly})--(\ref{eqn:d2q2qonly}), we have found 
(\ref{nls:WZ}), which is an NLS equation in 2+1 dimensions. 

\subsection{The elliptic NLS equation}
\label{subsubsec:ellipticity}

To make further progress towards understanding the form of 
possible solutions of the system (\ref{nls:WZ})--(\ref{Dacdef}), 
we make the substitution 
\beq
\zeta=\lambda Z, \qquad \xi = W - \frac{ D_M Z}{2D_Z} , 
\lbl{ZW-xizeta} \eeq 
to remove the mixed derivative term.  This transformation yields 
\beq
4 i w F_T = D_Z \left( \lambda^2 F_{\zeta\zeta} +  
\frac{(4 D_WD_Z - D_M^2)}{4 D_Z^2} F_{\xi\xi} \right) 
+ 3 b |F|^2 F (\beta C+\beta^*C^* - 6) . 
\lbl{eqn:standardfinalNLS} \eeq

The NLS equation in 2+1 dimensions has two forms 
depending on whether the second-differential operator part of 
the equation is elliptic or hyperbolic. We are only interested in 
elliptic systems  (where the coefficients of $F_{\xi\xi}$ and 
$F_{\zeta\zeta}$ have have the same sign) as our aim is to find 
solutions which are localised in both spatial dimensions. 
We therefore define the ellipticity as 
\beq
{\cal E}(k,l) = 4 D_W D_Z - D_M^2 , \qquad 
\lambda  = \frac{\sqrt{{\cal E}}}{2 D_Z} , 
\lbl{eqn:e} \eeq 
where expressions for $D_W$, $D_Z$ and $D_M$ are given in 
(\ref{Delta-def})--(\ref{Dacdef}).   Since we have two expressions for 
$C$, one for the acoustic mode and the other for the optical mode, 
as given in (\ref{eqn:C}), we have different expressions for $D_Z$, 
$D_W$, $D_M$ in the two cases, and two expressions for the 
ellipticity, ${\cal E}_{ac}$ and ${\cal E}_{opt}$. 

\begin{figure} [ht] 
{\centering 
\includegraphics[width=75mm]{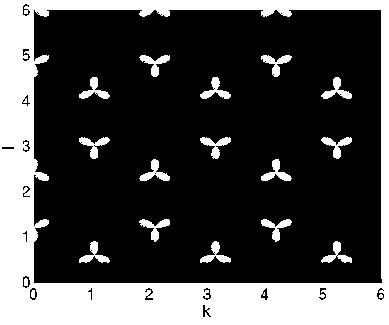} \  
\includegraphics[width=75mm]{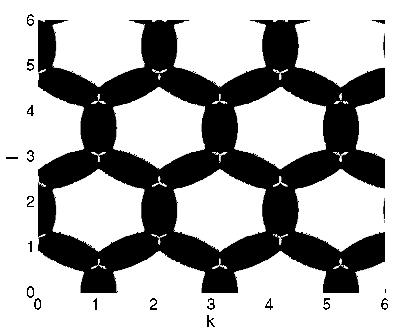} }   
\caption{Left: plots of the region where the function 
${\cal E}_{ac}(k,l)>0$, showing this to be negative almost 
everywhere (dark areas), only positive in small areas near 
the Dirac points (marked in white); 
right: plot of the region where ${\cal E}_{opt}(k,l)$ is positive (white), 
showing large areas, around maxima of the frequency $\omega_{opt}$, 
(eg $(k,l)=(0,0)$) and small areas near the Dirac points. }
\label{fig:e}
\end{figure}

In Figure \ref{fig:e} we plot the sign of the ellipticity functions 
${\cal E}(k,l)$ from (\ref{eqn:e}) for the acoustic and optical cases. 
In the acoustic case, ${\cal E}_{ac}\leq0$ for almost all $(k,l)$, 
there being small trefoil-shaped areas of positive ellipticity near 
the Dirac points.   However, for the optical case, there 
is a wide range of wavenumbers where ${\cal E}_{opt}>0$, as 
shown by the white hexagonal areas in the right panel of Figure 
\ref{fig:e}.  Note that the optical case also shows small 
trefoil-shaped areas of positive ellipticity near the Dirac points. 
Breathers corresponding to these wavenumbers are expected to be 
unstable as their frequencies will coincide with those of linear waves.  
Whilst the dispersion relation in this diatomic system does not have 
a gap -- the frequency spectrum $\omega(k,l)$) includes all values 
from zero to $\omega_{{\rm max}}=\sqrt{3}$, the form of the breathers 
near the Dirac points are expected to be similar to those of gap 
solitons in other diatomic systems, where the dispersion relation has 
gaps. One-dimensional FPU problems have been  
studied by Livi {\em et al}.\ \cite{Livi} and James \& Noble \cite{JN}. 
As in one-dimensional diatomic systems, the breathers corresponding 
to the optical domain including $(k,l)=(0,0)$ have frequencies which 
lie above the optical band, and so are expected to be long-lived. 

We now focus the optical case, where $C=C_{opt} = -\ee^{i\theta}$, 
and (\ref{eqn:standardfinalNLS}) simplifies to 
\beq
4 i \omega F_T = \lambda^2 D_Z \nabla^2_{(\xi,\zeta)} F 
- 6 b (3+|\beta|) |F|^2 F . 
\lbl{nls4} \eeq
In order for bright breathers to exist, there is a second criterion 
to be satisfied, namely that the coefficients of the 
nonlinear term and the spatial derivative must have the same sign. 
Since the nonlinearity is negative, we require $D_Z<0$. 
For the optical mode, this condition is satisfied for all $(k,l)$. 

\subsection{Asymptotic estimates for breather energy}
\label{subsec:estenergy}

The total electrical energy in the honeycomb lattice is conserved. 
This quantity is related to the Hamiltonian (\ref{eqn:ham}) by $E = 
\tilde{H}/C_0$. Thus, upto quadratic order, 
\beqa
H = C_0 E &=&  \half \sum_{m,n} \hat Q_{m+2,n}^2 + \ol Q_{m,n}^2 + 
(\ol P_{m,n}-\hat P_{m+2,n})^2 + (\ol P_{m,n}-\hat P_{m-1,n-1})^2 
\nn \\ && \qquad + (\ol P_{m,n}-\hat P_{m-1,n+1})^2 . 
\lbl{eqn:final-h-energy} \eeqa
Now our aim is to work out an expression for energy at leading order 
in $\ep$, given our solution for $\hat Q$, $\ol Q$ in terms of $F$. 
Since we are only interested in leading order approximation to the 
energy, the dependence of the solution for $F$ given by (\ref{Fsol}) 
on $T$ can be ignored, as the dependence on $\omega$ dominates. 
However, in passing we note that from (\ref{Fsol}) that the 
combined frequency of the breather mode is given by 
\beq
\Omega = \omega + \frac{3b \ep^2 A^2 (3+|\beta|)}{4\omega} , 
\lbl{Omega} \eeq
and so, in the optical case, the frequency lies above the frequency 
of linear waves. From 
(\ref{eqn:expanofqmnright})--(\ref{eqn:expanofqmnleft}) we find 
\beqa 
\ol Q_{m,n} &=& 2 \ep A \phi(r) \cos (km+lhn+\omega t ) , \nn\\  
\hat Q_{m,n} &=& - 2 \ep A \phi(r) \cos (km+lhn+\omega t + \theta ) , 
\lbl{Qsol} \eeqa  
where $r$ is given by the argument of $\phi$ in (\ref{Fsol}), and using 
$P=CF=-\ee^{i\theta}F$ since we are considering optical modes. For 
both left- and right-facing nodes, $\hat Q_{m,n}$ and $\ol Q_{m,n}$, we 
have ${\rm d}^2 Q/{\rm d}t^2 = - \omega^2 Q$, and ${\rm d} P/{\rm d}t = 
- Q$; hence, at leading order, ${\rm d} Q/{\rm d}t = \omega^2 P$ and 
\beqa 
\ol P_{m,n} &=& -\frac{2 \ep A}{\omega} \phi(r) \sin (km+lhn+\omega t),\nn\\
\hat P_{m,n} &=& \frac{2\ep A}{\omega}\phi(r) \sin(km+lhn+\omega t+\theta) . 
\eeqa 

The equation for the energy is given by (\ref{eqn:final-h-energy}), 
here we show the calculation of the onsite ($\hat Q_{m,n}$, 
$\ol Q_{m,n}$) part of this, the calculation of the interaction energy 
(due to $\ol P_{m,n}$, $\hat P_{m,n}$) can be found in a similar way 
and gives an identical final expression. 

We replace the double sum over $(m,n)$ in (\ref{eqn:final-h-energy}) 
by an integral over $(X,Y)$-space using (\ref{eqn:variables}), 
and transform into an integral over $(Z,W)$-space using (\ref{tw}). 
The Jacobian required to then make the transformation from 
$(Z,W)$ to $(\xi,\eta)$ coordinates using (\ref{ZW-xizeta}) is
$\left|\frac{\partial(\xi,\eta)}{\partial(Z,W)}\right| = \lambda$. Hence 
\beqa 
2 C_0 E &\!&\!  = \sum_{ {{m,n \ \rm{s.t.}}\atop{\ol Q_{m,n} \rm{exists}}}} 
\ol Q_{m,n}^2 + \hat Q_{m+2,n}^2 
= \int \int 4 A^2 \phi^2 \frac{\dd Z \dd W}{h} 
= \frac{ 2 \pi  I {\cal E} }{ 3 b h (3+|\beta|) } , 
\lbl{eqn:Energy1}\eeqa
where $I := \int_0^\infty r \phi^2(r) \, \dd r$. 
The final stages use $r=(A/\lambda)\sqrt{(\xi^2+\zeta^2) 3 b (3+|\beta|) 
/ (-D_z)}$ and (\ref{eqn:e}).  The above calculation makes use of 
the result $\sum_{m,n} \phi^2(r) \cos(km+lhn+\omega t)=0$ since $r$ is 
slowly varying in $m,n$. 

From equation (\ref{eqn:Energy1}), we note that to leading order, the 
energy of the breather is independent of the breather amplitude, $A$. 
Thus, no matter how small the breather amplitude, there is a minimum 
energy required to create it. The reason for this threshhold energy is 
that as the amplitude reduces, the width of the breather increases, in 
such a way that the total energy remains constant. This property 
confirms the observations of Flach \textit{et al}.\ in \cite{flach1}. 

\begin{figure}[ht] 
\begin{center}
\includegraphics[scale=0.6]{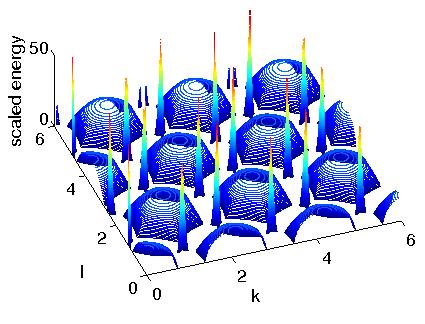}
\end{center}
\caption{Plot of $-\lambda D_Z / (3+|\beta|)$ against $(k,l)$, 
this being the ($k,l$)-dependent part of the breather energy 
$E(k,l)$ (\ref{eqn:Energy1}), (in colour in online version).} 
\label{fig:energysym}
\end{figure}

However, the threshold energy {\em is} dependent on the 
wavenumbers $k$ and $l$, therefore moving breathers have different 
threshold energies.   Figure \ref{fig:energysym} shows that the energy 
threshold is locally maximised at $k=l=0$, that is, for static breathers.  
Moving breathers require {\em less} energy to form.  An alternative 
viewpoint is that as breathers lose energy, they start moving, and 
accelerate, to the maximum speed, where the ellipticity constraint 
is only just satisfied.  It is also clear from (\ref{eqn:Energy1}) that 
the energy is closely related to the ellipticity constraint. 
Finally, we note that the breathers predicted near the Dirac 
points, which have frequencies lying in the linear spectrum,  
have much higher energies than the out-of phase optical breathers 
whose frequencies lie above the top of the linear spectrum. 

\section{Static breathers in an asymmetric potential}
\label{subsec:asym}

In this Section we examine the more general case for which 
$a\neq 0$ in (\ref{eqn:d2q1qonly}) and (\ref{eqn:d2q2qonly}).
Recall that at the end of Section \ref{sec:gen} we obtained a system 
of two coupled equations for $G_0$ and $F$, namely (\ref{goeq}) 
and the equation that can be derived by following the procedure 
(i)--(iv) in Section \ref{subsec:nls}.  It is only possible to reduce 
this system to a single solvable equation when $a=0$ (as analysed 
above in Section \ref{subsec:sym}) and when $G_0$ is independent 
of $\tau$, which we discuss in this section. 

The only example where the system becomes independent of $\tau$ 
is the case $k=l=0$, on the optical branch.  Under these conditions, 
we have, from (\ref{beta-def}), (\ref{eqn:dispersionrelation}), 
(\ref{eqn:C}), (\ref{betak-betal}), (\ref{eqn:u}), (\ref{eqn:v}), 
(\ref{tw}), (\ref{eqn:uhatvhat}) 
\beqa & 
\beta=3 , \quad \theta=0, \quad \omega_{opt} = \sqrt{6}, \quad 
C=-1, \quad \beta_k=\beta_l=0 , \quad u=v=c=0 , & \nn \\ &  
Z \equiv X, \quad W \equiv Y , \quad \hat u = \hat v = 0 .  
& \eeqa 
Since $u=v=0$, in this case, the breather is stationary; in addition, 
from (\ref{mat:g2q2}), we find ($\gamma=3$ and) $G_2=Q_2=0$. 
Assuming $G_0$ is independent of $\tau$,  equation (\ref{goeq}) 
can be solved by $G_0=Q_0=-a|F|^2$, enabling us to perform 
stage (ii) of the process outlined in Section \ref{subsec:nls}. 

Now we turn to deriving the NLS equation.  From stage (i) 
in Section \ref{subsec:nls}, and since $C=-1$, we form the 
equation $A=B$ from (\ref{Adef})--(\ref{Bdef}).  Since $P=-F$, 
stage (iii) of the calculation leads to 
\beq
2 i \sqrt{6} F_T + 3(F_{XX} + F_{YY} ) + 6 (3b-4a^2) |F|^2 F =0 . 
\lbl{nls5} \eeq
For bright breather solutions to exist, we require the coefficients 
of the nonlinearity and the spatial diffusion terms to have 
same signs, that is, $b > \mfrac{4}{3} a^2$.  
In place of (\ref{Omega}), the breather's frequency is now given by 
$\Omega=\sqrt{6} + 3\ep^2A^2 (3b-4a^2) / 2 \sqrt{6} $, which 
still lies above the top of the phonon band. 

\subsection{${\cal O}(\ep^3\ee^{3i\psi})$: expression for the third harmonic}
\label{H3-subsec}

As noted above, the second harmonic terms, $G_2$ and $Q_2$ 
are both zero for the case of stationary breathers, that is the 
optical mode with $k=l=0$. Hence we extend the expansion of 
Section \ref{sec:gen} to consider the terms at ${\cal O}(\ep^3 
\ee^{3i\psi})$ to see if third harmonic terms are generated. At 
${\cal O}(\ep^3\ee^{3i\psi})$, we obtain similar equations to those 
of Section \ref{subsec:e2epsilon2}, more specifically we obtain 
\beqa
-9\omega^2 H_3 & = & (\ee^{6ik} + \ee^{-3ik+3ilh} +\ee^{-3ik-3ilh}) 
( R_3 + b P^3 + 2 a  P Q_2) - 3 H_3 \nn \\ && - 3 b F^3  - 6 a FG_2 ,\nn\\
-9\omega^2 R_3 & = & (\ee^{-6ik} + \ee^{3ik-3ilh} + \ee^{3ik+3ilh}) 
( H_3 + b F^3  + 2 a F G_2 ) - 3 R_3 \nn \\ && - 3 b P^3 - 6 a PQ_2 . 
\eeqa 
These are the equations for general $k,l$; however, for stationary 
breathers we are only concerned with $k=l=0$, in which case 
$Q_2=G_2=0$, $\omega=\sqrt{6}$ and $P=-F$, hence we obtain 
the solution $H_3 = \rec{8} b F^3$, $R_3 = - \rec{8} b F^3$. 
Thus we find that the honeycomb lattice generates third harmonics 
but not second harmonics in the stationary breather. 

\subsection{Comparison with other lattice geometries}
\label{geom-comparison-subsec}

In earlier papers \cite{Buttwatt1,Buttwatt2} we have carried out 
similar calculations on the square and hexagonal lattices, where 
the derivations are considerably simpler.   In all cases we have 
$G_0=-a|F|^2$; however, other properties of the lattices differ, 
depending on the geometries concerned. In Table \ref{geom-tab} 
we compare the results of the honeycomb lattice analysed here 
with corresponding results for the square and hexagonal lattices 
analysed earlier.  

\begin{table}[!ht]
\begin{center}
\begin{tabular}{|c|c|c|c|} \hline 
Property \ $\backslash$ \ Geometry & 
Square \cite{Buttwatt1} & Hexagonal \cite{Buttwatt2} & Honeycomb \\[0.2ex] \hline 
Second harmonic & 
	$G_2=0$ & $G_2=\rec{3}a F^2$ & $G_2=Q_2=0$ \\[0.2ex]
Third harmonic & 
	$H_3=\rec{8} b F^3$ & $H_3=0$ & $H_3=-R_3=\rec{8} b F^3$ \\[0.2ex]
Inequality relating nonlin coeff & 
	$b>\mfrac{4}{3} a^2$ & $b>\frac{10}{9}a^2$ & $ b > \frac{4}{3} a^2$ \\[0.2ex] \hline 
\end{tabular}
\end{center}
\caption{Table summarising various properties of the 
different lattice geometries. }
\label{geom-tab}
\end{table}

The absence of any second harmonic is a property shared with 
the square lattice. Whilst the hexagonal lattice generates no third 
harmonic, it does generate a second harmonic.  Furthermore, the 
inequality relating the coefficients of nonlinear terms is identical for the 
honeycomb lattice and the square lattice, whilst different for the hexagonal. 
The possibly surprising result from this table is that, at least as far as 
stationary breathers are concerned, the honeycomb lattice has more in 
common with the square lattice than the hexagonal lattice.  

\subsection{Stability of the breather}
\label{subsec:stab}

The solution for $F$ is a one-parameter family, which 
we parametrise by the amplitude, $A$, as 
\beq
F = A \exp \left( \frac{3 i b A^2 T (3+|\beta|)}{4\omega} \right) \phi \left( 
\frac{A}{\lambda} \sqrt{\frac{3b(3+|\beta|)(\xi^2+\zeta^2)}{-D_Z}} \right) , 
\lbl{Fsol} \eeq
where $\phi(r)$ is the function which solves the elliptic problem 
$\nabla^2 \phi = \phi - \phi^3$in two dimensions.  This elliptic problem is 
known to have solutions, and the cylindrically-symmetric solution we write as 
$\phi(r)$. Solutions such as (\ref{Fsol}) are known as Townes soliton 
solutions \cite{chiao1} of the 2D NLS. These solutions are known to be 
unstable in the two-dimensional NLS, with subcritical initial conditions 
suffering from dispersion, leading to the amplitude converging to zero 
everywhere through the initial data spreading out; whilst supercritical 
initial conditions blow up, with the energy being focused to a single point, 
where the amplitude diverges.  However, arbitrarily small structural 
perturbations to (\ref{nls4}) can stabilize the Townes soliton. For example, 
results proven by Davydova {\em et al}.\ \cite{Davydova} for the equation 
\beq
i F_T + D \nabla^2 F + B |F|^2 F + P \nabla^4 F + K |F|^4 F = 0 ,
\lbl{davydova} \eeq
demonstrate the stability of a localised breather mode if $BD>0$ and 
$PK>0$. Clearly the presence of a higher derivative terms can mollify the 
blow-up singularity, whilst higher order nonlinearities can also help 
stabilise the soliton, as discussed by Kuznetsov \cite{Kuznetsov}. 
 
If we pursue higher order correction terms, for example, 
from ${\cal O}(\ep^5 \ee^{i\psi})$ terms, then terms such as $\nabla^4 F$ 
and $|F|^4F$ occur, which may stabilise the Townes soliton 
provided their coefficients have the correct combinations of signs. 
However, such an expansion also yields terms of the form 
$\nabla^2 (|F|^2 F)$, $|F|^2 \nabla^2 F$ and $F^2 \nabla^2 F^*$, and 
the effect of such structural perturbations of (\ref{nls4}) has, to our 
knowledge, not yet been determined.  Fibich and Papanicolaou 
\cite{fibich1,fibich2} have also addressed this problem, though their results 
do not yet extend to these nonlinear derivative terms.  

To illustrate this, let us consider the case of stationary breathers on a 
symmetric lattice, that is, we take the nearest-neighbour restoring force 
to be $V'(\phi) = \phi+b\phi^3+g\phi^5$ (that is, $a=0$ and no quartic 
nonlinearity).   We analyse the special case given by $k=l=0$, so that 
$u=v=0$, $G_0=Q_0=Q_2=G_2=0$.   Note that we also have $\hat u = 
\hat v=0$ so that $\hat G=0$, $H_0=R_0$ and we can take $\ol{G}=0$ 
so that $G_1=Q_1=0=H_0=R_0=0$.  Hence, in place of the ansatzes 
(\ref{eqn:expanofqmnright})--(\ref{eqn:expanofqmnleft}), used earlier, 
we use the simplified forms 
\beqa
\ol{Q}=\ep\ee^{i\psi}F+\ep^3\ee^{3i\psi}H_3+\ep^5\sum_{j=1}^5\ee^{ij\psi}J_j,
&\;\;\;&
\hat{Q}=\ep\ee^{i\psi}P+\ep^3\ee^{3i\psi}R_3+\ep^5\sum_{j=1}^5\ee^{ij\psi}U_j, 
\nn \\ && \eeqa
with $H_3 = \mfrac{1}{8} b F^3$, and $R_3 = 
-\mfrac{1}{8} b F^3$ as derived in Section \ref{H3-subsec}.   
Our aim is to calculate the form of the higher-order terms, namely those 
at ${\cal O}(\ep^4\ee^{i\psi})$ and ${\cal O}(\ep^4\ee^{i\psi})$.

Combining the results from ${\cal O}(\ep \ee^{i\psi})$, ${\cal O}(\ep^3
\ee^{i\psi})$, ${\cal O}(\ep^5 \ee^{i\psi})$, we obtain the governing equations 
\beqa
\lefteqn{ (3\!-\!\omega^2)  \ep F + \ep^5 J_1) - 3 ( \ep P + \ep^5 U_1)}
& & \nn \\ 
&=& - 2 i \omega \ep^3 F_T + 3 \ep^3 \nabla^2 P 
+ 9 b \ep^3 ( |P|^2 P - |F|^2 F ) - \ep^5 F_{TT}  \nn \\ && 
- 2 i \omega \ep^5 F_{\tilde T} + \ep^4 (P_{XXX}-3P_{XYY}) 
+ \mfrac{3}{4} \ep^5 \nabla^4 P 
+ 6 \ep^5 \nabla.\tilde{\nabla} P  \nn \\ && 
+ 3 b \ep^5 \nabla^2 (|P|^2P) 
+ \ep^5 ( 30 g + \mfrac{9}{8} b ) ( |P|^4P - |F|^4F) , 
\lbl{hoteq1} \\ 
\lefteqn{-3 (\ep F + \ep^5 J_1) + (3\!-\!\omega^2)( \ep P + \ep^5 U_1)}
& & \nn \\
& = & - 2 i \omega \ep^3 P_{T} + 3 \ep^3 \nabla^2 F 
+ 9 b \ep^3 9 ( |F|^2F - |P|^2 P ) - \ep^5 P_{TT} \nn \\ &&  
- 2 i \ep^5 \omega P_{\tilde T} +  \ep^4 (-F_{XXX}+3F_{XYY}) 
+ \mfrac{3}{4} \ep^5 \nabla^4 F 
+ 6 \ep^5 \nabla.\tilde{\nabla} F \nn \\ && 
+ 3 b \ep^5 \nabla^2 (|F|^2F) 
+ \ep^5 ( 30 g + \mfrac{9}{8} b ) ( |F|^4 F - |P|^4 P) . 
\lbl{hoteq2} \eeqa
Here, in addition to the long scales defined in (\ref{eqn:variables}), we 
have introduced even longer time and length scales given by 
$\tilde T = \ep^4 t$, $\tilde X = \ep^3 m$, and $\tilde Y = \ep^3 h m$, 
and $\tilde{\nabla}$ is the corresponding vector derivative with 
respect to $\tilde X$ and $\tilde Y$. 

Since $\omega=\sqrt{6}$, the right-hand-sides of (\ref{hoteq1})--(\ref{hoteq2}) 
must be equal.  Combining this with the relation $P=-F$ leads to 
\beqa
0 & = & 2 i \sqrt{6} (F_T + \ep^2 F_{\tilde T}) + 3 \nabla^2 F + 18 b |F|^2 F 
\lbl{honls}\\ && + \ep^2 \left[ F_{TT} + \mfrac{3}{4} \nabla^4 F + 
6 \nabla.\tilde{\nabla} F + 3 b \nabla^2 (|F|^2F) + (60 g +
\mfrac{9}{4} b^2) |F|^4F\right] , \nn  
\eeqa
since the third-derivative terms cancel. Whilst these terms generate 
nonzero solutions for $J_1,U_1$, such contributions do not concern 
us here, where our aim is to determine the properties of $F$.  
The effect of the $F_{\tilde T}$ term is to change the timescale slightly, 
and the $\nabla.\tilde{\nabla}F$ term rescales the space scale $X$, 
hence we will neglect these terms. 

Applying $2i\omega \partial_T$ to the leading order form of 
(\ref{honls}), which is (\ref{nls5}) in the case $a=0$, yields 
\beq
0 = 8 F_{TT} + 3 \nabla^4 F + 108 b^2 |F|^4 F + 18 b ( \nabla^2 ( |F|^2 F) 
+ 2 |F|^2 \nabla^2 F -  F^2 \nabla^2 F^* ) , 
\eeq
which we use to eliminate $F_{TT}$ from (\ref{honls}), to find the 
final governing equation 
\beqa
0 & = & 2 i \sqrt{6} F_{T} + 3 \nabla^2 F + 18 b |F|^2 F 
+ \mfrac{3}{8} \ep^2 \nabla^4 F 
+ (60 g - \mfrac{45}{4} b^2) \ep^2 |F|^4 F  \nn \\ && 
+ \mfrac{3}{4} b \ep^2 \nabla^2 (|F|^2F)  
+ \mfrac{9}{4} b \ep^2 F^2 \nabla^2 F^* 
- \mfrac{9}{2} b \ep^2 |F|^2 \nabla^2 F . 
\eeqa
As the last three terms do not appear in (\ref{davydova}), we cannot formally 
determine the stability properties of the system.  However, if we were 
to simply ignore the last three terms, (\ref{davydova}) suggests that if 
$g > \mfrac{3}{16} b^2$ (and $b>0$) then the combined influence
of the fifth-order nonlinearities and fourth order derivatives stabilise the 
breather.   Since we expect that the second derivatives of cubic 
nonlinearities can be bound by some combination of fifth order 
nonlinearities and fourth order derivatives of $F$, it is reasonable to 
assume that for sufficiently large $g$, the breather will be stable. 

\section{Conclusions}
\label{sec:conc}

We have investigated the properties of discrete breathers on a 
two-dimensional honeycomb lattice. After applying Kirchoff's laws 
to the electrical lattice, we derived a governing set of equations for 
the case of nonlinear capacitors at nodes, and nodes being connected 
by linear inductors. Using multiple scales asymptotic methods, we 
reduced the governing equations to a single NLS equation from which 
we can determine the properties and conditions under which 
small-amplitude breathers may exist.  There are two cases in which an 
NLS equation can be obtained. We analysed each case in more detail 
in Sections \ref{subsec:sym} and \ref{subsec:asym}. 

The analysis of the honeycomb is more complicated than either the 
square or the triangular lattices, due to the necessity of treating the 
two types of node, which means that a diatomic analysis must be 
carried out. This leads to extra complications at the level of 
determining the evolution of the `slow mode' at ${\cal O}(\ep^4 \ee^0)$.  
Part of the extra complexity is that in order to derive the leading 
order $F$ and $P$ terms in (\ref{eqn:expanofqmnright}) and 
(\ref{eqn:expanofqmnleft}), it is necessary to simultaneously 
find the first correction terms $G_1$ and $Q_1$. 

The first special case we considered (\S\ref{subsec:sym}) was that 
of a symmetric potential in which the terms $G_0$, $Q_0$, $Q_2$ 
and $G_2$ are all zero.   From this we were able to obtain an ellipticity 
condition, for the wavenumbers $(k,l)$, to ensure we obtained solutions 
which were localised in both spatial directions.  A minimum threshold 
energy to create breathers was also found. This confirmed the 
observations of Flach {\em et al.}\ \cite{flach1}.  The ellipticity condition, 
breather energy and dispersion relation, obtained in Sections 
\ref{subsec:disp} and \ref{subsec:sym}, were plotted. The breather energy 
is maximised for these stationary breathers. For other wave vectors, 
moving breathers are created, with lower energies.

The second case we analysed was the case of asymmetric potentials. 
Here we only considered the specific wavenumber $k=l=0$, and the 
optical branch which guarantees stationary breathers. However, this 
enabled us to describe the behaviour of a range of nonlinearity 
parameters ($a,b$) for which stationary breathers may exist. We find 
no second harmonic term in the expansion in this case, but there is a 
third harmonic. These properties show a close similarity between the 
square lattice and the honeycomb, quite distinct from the hexagonal lattice. 

It is natural to consider the relationship between the honeycomb 
system studied here and a one-dimensional systems.  
We note that the two-dimensional systems studied previously 
\cite{Buttwatt1,Buttwatt2} both had a dispersion relation with a 
single branch that described modes with optical and acoustic 
characters, as in one-dimensional (monatomic) systems.  However, 
in one-dimensional {\em diatomic} systems,  the dispersion relation 
has two branches, one optical and one acoustic.   Such systems 
have been studied by Livi {\em et al}.\ \cite{Livi} and James \& 
Noble \cite{JN}, amongst others. In the latter paper, the authors 
derive the dispersion relation, showing it to have two branches: 
an optical and an acoustic form, as in the honeycomb lattice.  
However, in the one-dimensional diatomic lattice, the two branches 
are distinct, and do not meet at Dirac points; rather, there is a gap 
between the two branches, in which breathers may exist with 
frequencies which are not coincident with any linear wave.  
In the honeycomb lattice, the two branches meet at the Dirac points, 
and so there is no gap.   However, from Figure \ref{fig:e} we see that 
the honeycomb lattice still supports breather solutions near the Dirac 
points. Whilst it would be interesting to investigate these solutions 
further, we expect them to be unstable, since their frequencies 
coincide with those of linear waves, allowing energy interchange with 
phonons which could lead to the breather's decay.  In contrast, we 
now consider the larger white regions in Figure \ref{fig:e} (right), 
corresponding to wavenumbers for which optical breathers exist. 
We expect these breathers to have frequencies above the top of the 
optical band, and so will have no linear wave with the same frequency.  
These waves, however, may still be unstable, due to other effects, 
such as the breather's motion over lattice sites being resonant with 
linear modes.  Such losses may still allow the breather to travel long 
distances before decaying, and so be relevant in applications such 
as the explanation of tracks in mica via quodons as suggested by 
Russell and Eilbeck \cite{mike,russel1}.

In this paper we have only looked at a scalar-valued quantities at 
each node that is, only one degree of freedom.   In future works  
we aim to analyse the stability of these breather solutions, and find 
approximate solutions to the vector-valued honeycomb lattice 
similar to the lattices Marin \textit{et al.}\ studied in \cite{marin1}.

\subsection*{Acknowledgements}

We are grateful to Hadi Susanto for useful conversations. 


\end{document}